\documentclass{iopart}
\usepackage{iopams}
\usepackage{amssymb}
\usepackage{graphicx}
\usepackage{dcolumn}
\usepackage{bm}

\def\be{\begin{equation}}
\def\ee{\end{equation}}
\def\bea{\begin{eqnarray}}
\def\eea{\end{eqnarray}}

\def\T{\theta}
\def\t{\tau}
\def\s{\sigma}
\def\e{\varepsilon}
\def\ep{\epsilon}
\def\G{\Gamma}

\newcommand{\vfi}{\varphi }
\newcommand\D{\delta}
\newcommand{\Sc}{Schr\"odinger }
\renewcommand{\o}{\omega}

\begin{document}

\title{%
Dynamical qubit controlling via
pseudo-supersymmetry of two-level systems%
}

\author{Boris F. Samsonov and V. V. Shamshutdinova}
\address{Tomsk State University, 36 Lenin Avenue, 634050 Tomsk, Russia}
\date{\today}

\begin{abstract}
For a flux qubit considered as a two-level system, for which a
hidden polynomial pseudo-supersymmetry was previously discovered,
we propose a special time-dependent external control field. We
show that for a qubit placed in this field there exists a critical
value of tunnel frequency. When the tunnel frequency is close
enough to its critical value, the external field frequency may be
tuned in a way to keep the probability to detect a definite
direction of the current circulating in a Josephson-junction
circuit above 1/2 during a desirable  time interval. We also show
that such a behavior is not much affected by a sufficiently small
dissipation.
\end{abstract}

\maketitle

\section{Introduction}

In the past few years superconducting circuits based on Josephson
tunnel junction have attracted much attention both from
theoretical and experimental viewpoints as possible candidates for
the implementation of quantum computer (see, e.g. Refs.
\cite{FromEPJB34(269)1,FromEPJB34(269)2,fluxQbit}). Usually they
represent a small Josephson-junction circuit, called a Cooper-pair
box, which consists of a small superconducting electrode connected
to a reservoir via a Josephson junction \cite{FromEPJB34(269)2}.
For a flux qubit the circuit with a very small inductance
containing three Josephson junctions is described (in appropriate
units) by the following two-level Hamiltonian \cite{fluxQbit}:
\be\label{Hq} H_q=-\Delta\s_x-\varepsilon(t)\s_z\,. \ee Here
$\Delta$ is the tunnel frequency, and $\e(t)$ is a time-dependent
field (bias) which is controlled by an externally applied flux.
Although in general $\Delta$ is a function of $\varepsilon$, it
varies on the scale of the single junction plasma frequency, which
is much above the typical energy range at which the qubit is
operated \cite{D}. We thus can assume $\Delta$ to be constant for
the purpose of this paper. Solving the \Sc equation with
Hamiltonian (\ref{Hq}),
\begin{equation}\label{Scequation}
  i\dot\Psi(t)=H_q\Psi,\quad\Psi=(\psi_1,\psi_2)^T
\end{equation}
(superscript ``\emph{T}" means the transposition and the dot over
a symbol means the derivative with respect to time), we obtain the
probability of a definite direction of the current circulating in
the ring, i.e. $P^{\downarrow}=|\psi_{1}|^2$ is the probability of
the clockwise direction of the current and
$P^{\uparrow}=|\psi_{2}|^2$ is the probability of the opposite
current direction.

One of the most important problem in quantum computation is
connected with the possibility of controlling the state of an
array of qubits. Typically the simplest two-qubit operations are
generated by interplay of the coupling between qubits and local
fields. Much theoretical attention has been recently paid to
studying the controllable coupling between qubits of different
types (see Ref. \cite{q-control} and references therein). Recently
it has been shown \cite{Zhang} that in the simplest and the most
important from engineering viewpoint case of an ``always on" and
fixed coupling, a two-qubit Hamiltonian may be decoupled and the
control problem is, in particular, reduced to finding the
evolution of a one qubit placed in a time-dependent external
control field.  This observation shows an additional importance of
controlling a one qubit state. This is the main subject we devote
the present paper.

Usually, the probabilities, as functions of time, show an
oscillating behavior (cf. famous Rabi oscillations, see e.g. Ref.
\cite{Orszag}). But for some specific external fields this
character may be changed drastically thus showing a possibility to
control the qubit state. Up to now such a possibility is known to
be mostly related to oscillating external fields \cite{Drast}.

Recently \cite{Bagrov} it was proposed to consider a two-level \Sc
equation as a Dirac equation with a non-Hermitian Hamiltonian
where time plays the role of a spatial variable. This possibility
revealed a hidden pseudo-supersymmetry which may be associated
with a two-level system \cite{Shamshutdinova} and lead to
discovering new time-dependent external fields where a two-level
system admits solutions in terms of elementary functions. An
advantage of analytic solutions is the possibility of a careful
analysis of their properties which may reveal unexpected
peculiarities \cite{preprint}. In this paper we apply these
results to show the possibility of controlling the qubit state
with an external field of a special configuration. We show that
there exists a critical value of the tunnel frequency $\Delta$.
While the tunnel frequency approaches this critical value, the
probability $P^{\downarrow}(t)$ oscillates around a value
exceeding $1/2$ with a decreasing  amplitude and after the
critical value is reached it becomes a function monotonously
increasing up to limiting value equal $3/4$. Then using the
property that this special excitation regime is, in fact, a
limiting case of a more general oscillating external field, we
demonstrate that one can control a definite direction of the
current in the ring (i.e. the qubit state) during a desirable time
interval. Finally we also show that such a  behavior is not much
affected by the presence of a reasonably small dissipation
featuring open quantum systems. We start with reminding the reader
main constructions leading to polynomial pseudo-supersymmetry in
two-level systems which is done in the next section.

\section{Polynomial pseudo-supersymmetry of two-level systems}

Similar to the conventional supersymmetry in quantum mechanics
(for recent developments see a special issue of 2004 J. Phys. A:
Math. Gen. 34 (43)), pseudo-supersymmetry is based on intertwining
and factorization relations. However, in this case Hamiltonians
are non-Hermitian and a specific automorphism, that defines
pseudo-adjoint operators, should be involved \cite{Znojil,Mostafazadeh}.
In particular if $A$ is a linear operator and $\eta$ is linear,
Hermitian, invertible operator then
\begin{equation}
    A^\sharp=\eta^{-1}A^+\eta\,,
\end{equation}
where `$+$' sign denotes the usual
(e.g. Laplace)
 adjoint operation,
 by definition is
the pseudo-adjoint
of $A$ with respect to $\eta$.
The operation of
formal (Laplace) conjugation obeys the standard rules $(AB)^{\dag }=B^{\dag
}A^{\dag }$, $(d/dt)^{\dag }=-d/dt$ and corresponds to the transposition of
a matrix accompanied by the complex conjugation of its elements.
Operator $B$ is said to be pseudo-Hermitian with respect to $\eta$
if $B^\sharp=B$, i.e.
\begin{equation}\label{}
B=\eta^{-1}B^+\eta.
\end{equation}
Basic properties of pseudo-Hermitian operators are discussed in
detail in \cite{Mostafazadeh}.

Recently \cite{Shamshutdinova} it was observed that a
polynomial
pseudo-supersymmetry may be associated with a two-level
system
interacting with a classical (i.e. not quantized) electromagnetic
field.
The method is based on the possibility to rewrite the
Schr\"{o}dinger equation (\ref{Scequation}) that governs the
evolution of the system in the form of a one-dimensional
stationary Dirac equation
\begin{equation}\label{Dirac}
    h_0\Psi=E\Psi,
\end{equation}
where the time plays the role of a spatial variable and
the Dirac Hamiltonian
\begin{equation}\label{h0}
    h_0=i\sigma_x\frac{d}{dt}+V_0(t),\quad
    V_0(t)=i\sigma_y\varepsilon_0(t),\quad E=\Delta
\end{equation}
is non-Hermitian,
with the subsequent application of the
well-developed  intertwining
operators technique \cite{14}.
Function $\varepsilon(t)$ plays the role of a ``potential".

It is easy to see that $h_0$ (\ref{h0}) is pseudo-Hermitian with
respect to $\eta=\sigma_x$.
The next ingredient of the method is based on the existence
for any real-valued function $\varepsilon_0(t)$
 such real-valued function
 $\varepsilon_1(t)$
and operator $L$ (intertwiner) that
\begin{equation}\label{inter}
 Lh_0=h_1L
\end{equation}
 where
$h_1=i\sigma_xd/dt+V_1(t)$
with $V_1=i\sigma_y\varepsilon_1(t)$.
The pseudo-Hermiticity of $h_0$ results in the following
factrorizations
\begin{equation}\label{factor}
\eta L^{+}\eta L=h_{0}^{2}-\Lambda^{2}\,,\quad
L\eta L^{+}\eta =h_{1}^{2}-\Lambda^{2}\,.  \label{JLJ}
\end{equation}%
The constant matrix $\Lambda=\mathrm{diag}(\lambda,-\lambda)$
in (\ref{JLJ}) is called the (matrix) factorization constant ($\lambda$
is also called the factorization constant). Formulas (\ref{JLJ}) present a
generalization of the factorization properties of transformation operators
that take place in the case of Hermitian one-component Hamiltonians
\cite{Mielnik}.
If now we introduce matrix operators
(in block-matrix forms)
\be
H=\left(
\begin{array}{cc}
h_0 & 0\\ 0 & h_1
\end{array}
\right),  \quad Q_1=\left(
\begin{array}{cc}
0 & 0\\ L & 0
\end{array}
\right),  \quad Q_2=\left(
\begin{array}{cc}
0 & JL^+J\\ 0 & 0
\end{array}
\right) \ee then the intertwining (\ref{inter}) and factorization
(\ref{factor}) relations may be rewritten as the following set of
commutation and anti-commutation relations between these
operators: \be\fl Q_{1,2}^2=0\,,\quad HQ_{1,2}=Q_{1,2}H\,,\quad
Q_1Q_2+Q_2Q_1= H^{2}-\mathrm{diag}(\Lambda,\Lambda)\ee which
indicate on the simplest quadratic pseudo-superlagebra. Note that
the subsequent application of this technique leads to a more
general polynomial pseudo-supersymmetry \cite{Shamshutdinova}. On
the other hand if we start from $V_0$ with known solutions to the
Dirac equation (\ref{Dirac}) then solutions of the same equation
with $V_1$ can be obtained by applying $L$ to the previous
solutions. In this way new exactly solvable two-level potentials
are obtained \cite{Shamshutdinova} which we use in the next
section to demonstrate the possibility of the dynamical qubit
controlling.

\section{Dynamical qubit controlling}

\begin{figure}[t]
\begin{minipage}{13cm}
\includegraphics[width=6cm]{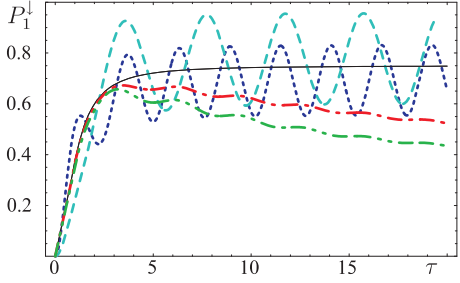}
\caption{(Color online) Evolution of $P_1^{\downarrow}$
probability at $\D=\sqrt3/2$ solid, dot-dashed and
double-dot-dashed (black, red and green) lines, at dephasing
($\Gamma _{\varphi}$) and relaxation ($\Gamma_{r}$) rates:
$\G_r=\G_\vfi=0.05$ dot-dashed (red) line and $\G_r=\G_\vfi=0.1$
double-dot-dashed (green) line; $\D=\sqrt3/2\pm0.25$ dotted and
dashed (violet and blue) lines resp.} \label{fig1}
\end{minipage}
\end{figure}

Consider first the case when the external control field
$\e=\e_1(t)$ changes in the following way:
\begin{equation}  \label{e-om}
\e_{1}\left(t\right) =-\e_0+\frac{4\e_0}{1+4\e_0^2t^{2}}\,.
\end{equation}%
Parameter $\e_0$ gives us the possibility to choose a suitable
time scale since after re-scaling $\t= 2\e_0t$, and redefining
parameter $\Delta$, $\Delta=2\e_0\delta$, we obtain the \Sc
equation with the Hamiltonian \be\label{e1} H=-\D\sigma_x
-\ep(\t)\sigma_z\,,\quad
\ep(\t)=\ep_1(\t)=-\frac12+\frac{2}{1+\t^2} \ee for which exact
analytic solutions are known \cite{Shamshutdinova}. Therefore
imposing the initial condition $P^{\downarrow}\left(0\right)=0$ we
can write down an explicit expression for the probability
$P^{\downarrow}\left(\t\right)$:
\begin{eqnarray}\nonumber
 P_{1}^{{\downarrow} }\left( \t \right) =
\displaystyle{
\frac{(\T^2-1)(\T^2+4)}{2\T^4}\frac{\t^2}{1+\t^2}+
\frac{(\T^2-1)(\T^2-4)}{2\T^6(1+\t^2)}
}
\\
\times
\left[
\T^2-4-(\T^2-4+\T^2\t^2)\,\cos\T\t+4\T\t\sin\T\t
\right]\label{P1}
\end{eqnarray}
where we have introduced $\T =\sqrt{1+4\D^{2}}$. From here it is
clearly seen that $P_{1}^{{\downarrow} }(\tau )$ is an oscillating
function provided $\T\ne 2$ ($\D \ne\sqrt 3/2$). For $\T= 2$
($\D=\sqrt3/2$)  equation  (\ref{P1}) yields
\begin{equation}  \label{P_mon}
P_{1}^{\downarrow}\left(\tau \right)=
\frac{3}{4}\frac{\t^2}{1+\t^2}
\end{equation}
which is a function monotonously increasing from zero at $\tau=0$
till the value $3/4$ for $\t \gg 1$ (solid (black) line in Fig.
\ref{fig1}). Since $\D$ differs from $\Delta$ only by the scaling
factor $2\e_0$ we will call $\D$ tunnel frequency as well. We note
a decrease of the oscillation amplitude when $\D$ approaches its
critical value equal $\sqrt{3}/2$. This is why for $\D$ close
enough to the value $\sqrt{3}/2$ the minimal value of the
probability $P_{1}^{{\downarrow} }(\t)$ for $\t>2$ exceeds $1/2$
(see dashed (blue) and  dotted (purple) lines in Fig. \ref{fig1}).

\begin{figure}[t]
\begin{minipage}{13cm}
\includegraphics[width=6cm]{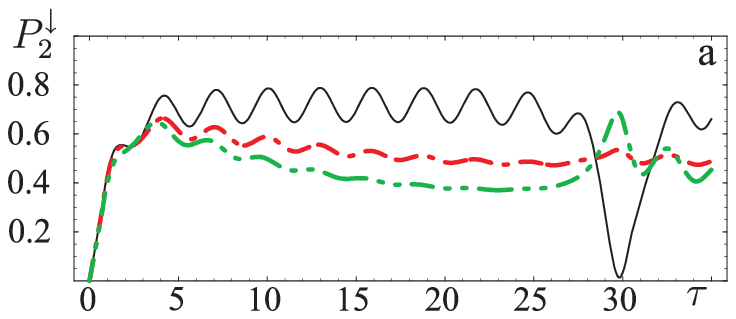}
\includegraphics[width=6cm]{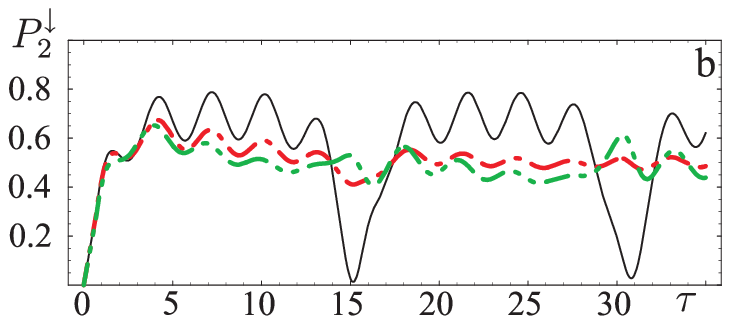}
\includegraphics[width=6cm]{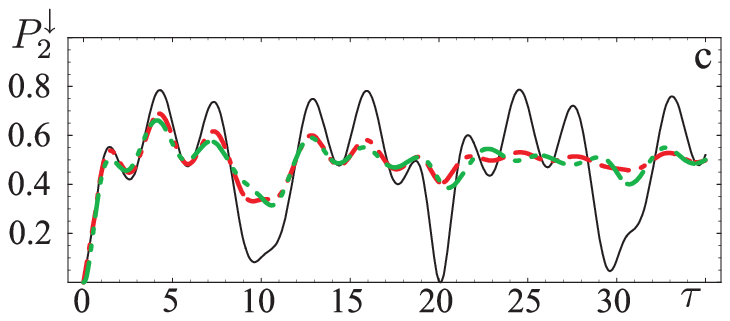}
\caption{(Color online)(a, c, d) Evolution of $P_{2}^{{\downarrow}
}$ probability at $\D=\sqrt{3}/2+0.1$, at dephasing ($\Gamma
_{\varphi}$) and relaxation ($\Gamma_{r}$) rates: $\G_\vfi=0.1$,
$\G_r=0.05$ dot-dashed (red) line and $\G_r=0.2$ double-dot-dashed
(green) line; (a) $\o=0.105$, (b) $\o=0.205$ and (c) $\o=0.314$.}
\label{fig2a} \label{fig2b} \label{fig2c}
\end{minipage}
\end{figure}

\begin{figure}[t]
\begin{minipage}{13cm}
\includegraphics[width=6cm]{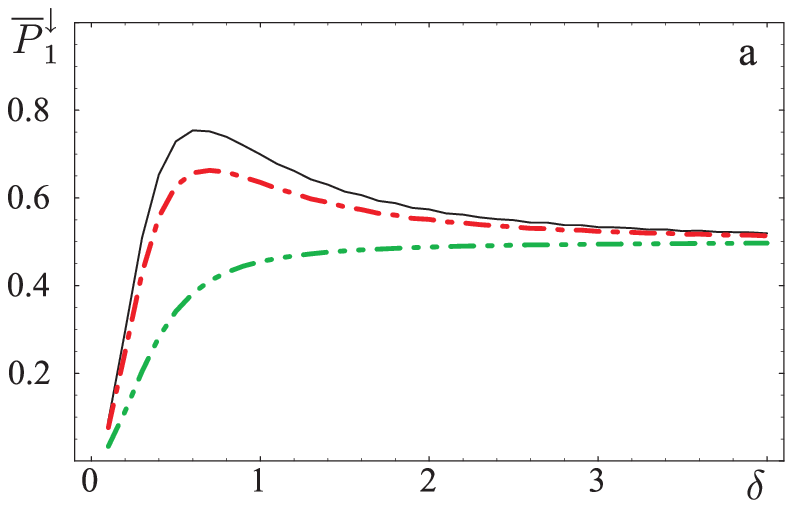}
\hspace{1cm}
\includegraphics[width=6cm]{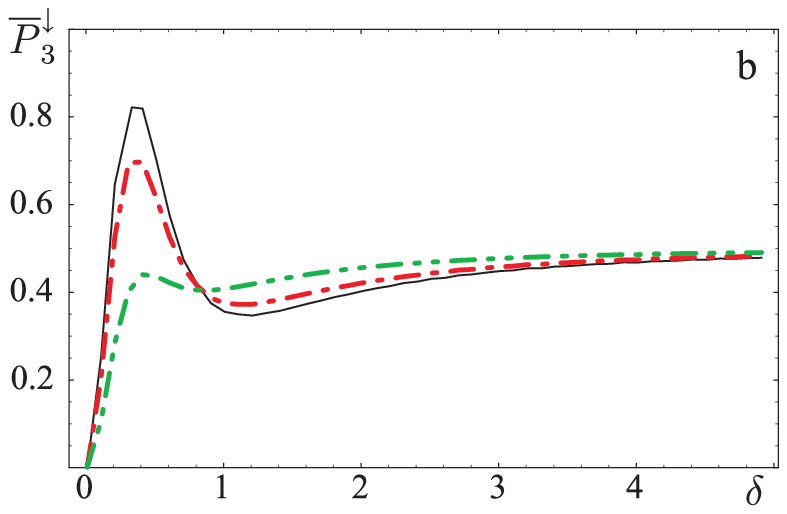}
  \caption{(Color online)
  $\D$-dependence of time-averaged probabilities for the
  closed system solid (black) lines and with dephasing ($\Gamma_{\varphi}$) and relaxation ($\Gamma_{r}$) rates:
 $\G_r=\G_\vfi=0.01$ dot-dashed (red) lines;
  (a)  $\G_r=\G_\vfi=0.1$ and (b)  $\G_r=\G_\vfi=0.05$
  double-dot-dashed (green) lines.}
\label{fig3a}\label{fig3b}
 \end{minipage}
\end{figure}

The time-averaged probability as shown in Fig. \ref{fig3a}a by
solid (black) line exhibits a maximum. Its analytic expression
\begin{equation}  \label{P1aver1}
\overline{P_{1}^{\downarrow}}=
2\D^2\frac{5+4\D^2}{\left( 1+4\D^{2}\right)^{2}}
\end{equation}
allows us to get the exact position of the maximum which is
$25/32\approx0.78$ at $\D=\sqrt{5/12}$.

The result we have just obtained suggests us to consider a more
general case \cite{Bagrov,Shamshutdinova}, where function
$\ep(\t)$ being periodical depends on three parameters. One of
them we fix by re-scaling both time and another parameter
(frequency $\omega $ appearing in Eq. (\ref{e3})) in a way as it
has been done above, thus obtaining the field
\begin{equation}  \label{e3}
\ep_{2}\left(\t\right) =
-1/2-\frac{2\omega ^{2}}{b\cos\left(2\omega\t+\varphi \right)
-1/2}
\end{equation}%
where $b^{2}=1/4-\o^{2}>0$. The remaining parameter $\varphi$ can
also be eliminated by shifting the time origin so that we put
$\varphi=0$ thus reducing external fields to a one parameter
($\o$) family. It is important to note that the previous result
(\ref{e1}) may be obtained from here at
$\varphi=\arctan\omega-\frac{1}{2}\arctan\frac{\omega}{b}$ in the
limit $\o\to0$.

According to \cite{Shamshutdinova} the
analytic expression for $P_{2}^{{\downarrow} }(\tau)$ reads
\begin{eqnarray}\nonumber
P_{2}^{\downarrow}\left(\t\right) & = &
 \frac{4\D^2}{\T^2}\sin^{2}
\left({\textstyle{\frac 12}} {\T\t}
\right)
\\
 & - &\frac{4\D^2b\left[
Q-\o(b+b^2-\D^2)\T\sin(2\o\t)\sin(\T\t)
\right]}{\T^2(b^2+\D^2)^2(2b\cos(2\o\t)-1)}
\nonumber
\end{eqnarray}
where \bea\nonumber Q & = &
b(1+2b)\T^2\cos^2\left({\textstyle{\frac 12}} \T\t \right)
\sin^2 \left(\o\t \right) \\
&+ &4\o^2(b-2\D^2)\cos^2\left(\o\t\right)
\sin^2\left({\textstyle{\frac 12}} {\T\t}\right).
\nonumber
\eea

To show the possibility to control the qubit state by external
field (\ref{e3}) we plot function $P_{2}^{\downarrow}(\t)$ for
$\D$ close to its critical value and for different values of $\o$.
Since the above considered case corresponds to $\o=0$, we show in
Fig. \ref{fig2a}a (solid (black) line) its behavior for $\o=0.105$
which is rather close to zero. During a sufficiently long time
interval the probability oscillates between $0.6$ and $0.8$ after
which it falls to zero. The closer gets $\o$ to zero, the longer
this period becomes and the closer to $1/2$ $\o$ becomes, the more
this interval is reduced (see the solid (black) lines in Figs.
\ref{fig2a}a, \ref{fig2b}b and \ref{fig2c}c). We have to note that
in the limit $\o\to1/2$ ($b\to0$ in Eq. (\ref{e3})) function
$\ep_2(\t)$ tends to a constant value equal $1/2$. This signal
reproduces the Rabi oscillations with the frequency
$2\sqrt{\D^2+1/4}$. Thus, $\o$ may be considered as a continuously
tunable parameter of the external field (\ref{e3}) with the help
of which, starting with the usual Rabi oscillations, one may   fix
the clockwise current direction as long as desirable. We would
like to stress that the range of $\D$, for which the probability
exceeds the value $1/2$, is rather large, i.e. $0.6<\D<1.1$. This
may facilitate its experimental detection.

\begin{figure}[tbp]
\begin{minipage}{13cm}
\includegraphics[width=6cm]{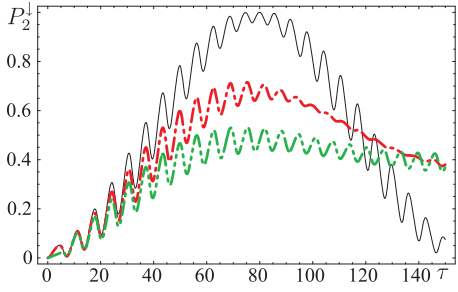}
\caption{(Color online) Time dependence of clockwise current
direction probabilities at $\o=0.49$ and $\D=0.1$ solid (black
line), at dephasing ($\Gamma _{\varphi}$) and relaxation
($\Gamma_{r}$) rates: $\G_r=\G_\vfi=0.01$ dot-dashed (red) line
and $\G_r=\G_\vfi=0.02$ double-dot-dashed (green) line.}
\label{fig4}
\end{minipage}
\end{figure}

Analytic expression for $P_{2}^{\downarrow}\left(\t\right)$ shows
us that it represents a complicated superposition of two
oscillating functions with frequencies $\T$ and $2\o$. Therefore
if these frequencies are close enough to each other one may
observe a beating phenomenon (Fig. \ref{fig4}). In this case the
oscillations with the small amplitude and the frequency close to
the Rabi frequency take place at the background of the
oscillations with the amplitude close to 1 and the very small
frequency defined by the difference between $2\o$ and $\T$.

Another aspect we would like to emphasize is that this type of the
external field is not unique. As we show below, there exist other
possibilities for the time dependence of the external field
exhibiting a similar feature.

\begin{figure}[tbp]
\begin{minipage}{13cm}
  \includegraphics[width=6cm]{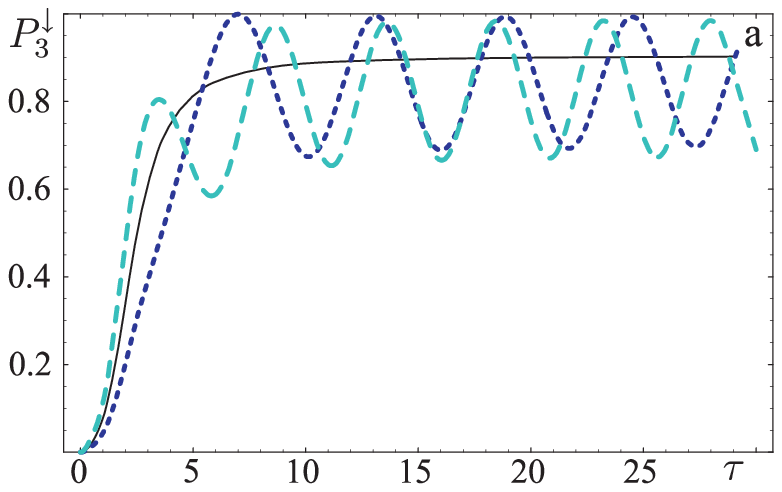}
  \hspace{1cm}
  \includegraphics[width=6cm]{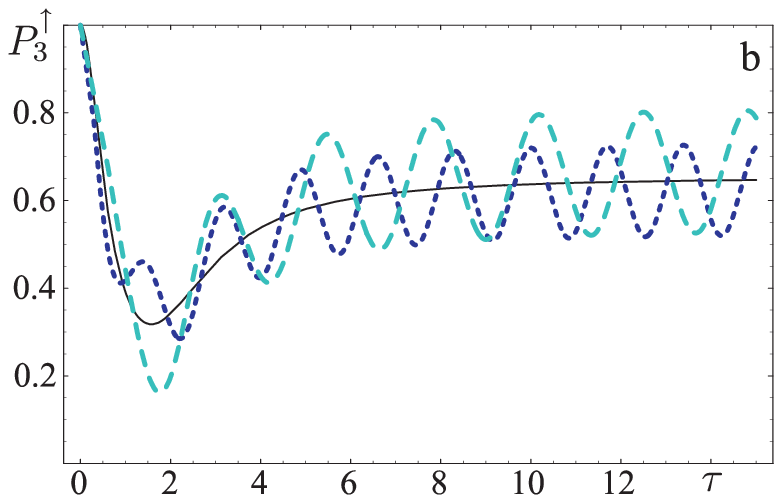}
  \caption{(Color online)
 Probabilities (a) $P_3^\downarrow(\t)$
at $\T=\sqrt 5-1$ solid (black) line,
 $\T=\sqrt5-1.1$ dotted (violet) line
 and  $\T=\sqrt 5-0.9$ dashed (blue) line;
  (b)  $P_3^\uparrow(\t)$ at $\T=\sqrt 5+1$ solid (black) line,
$\T=\sqrt5+1.5$ dotted (violet) line and $\T=\sqrt5+0.5$ dashed
(blue) line.}
 \label{fig5a}\label{fig5b}
 \end{minipage}
\end{figure}

Consider an exactly solvable model with a bit more
complicated form of function $\ep(\t)=\ep_3(\t)$ \cite{Shamshutdinova}
\begin{equation}\label{e2}
\ep _{3}\left( \t\right) =
-\frac12\,+
\frac{6}{Q_0}\left(\t^4+6\t^2-3\right)
\end{equation}%
where  $Q_0={\t^6+3\t^4+27\t^2+9}$.
The clockwise current direction probability
$P^{{\downarrow} }(\t)=P_3^{{\downarrow} }(\t)$
 in this case reads
\begin{eqnarray}\nonumber
P_{3}^{{\downarrow} }\left( \tau \right) & = &
\frac{4(\T^2-1)\t^2}{\T^8Q_0}
\left[144(1+\t^2)+\T^4(\t^2+9)^2-24\T^2(5\t^2+9)\right]
 \\
& + &
\frac{(\T^2-1)Q_1}{\T^{10}Q_0}
\left[Q_2\sin^2(\textstyle{\frac12}\t\T)+Q_3\sin(\t\T)\right]
\label{P2e2}
\end{eqnarray}%
where
\begin{eqnarray*}
Q_{1}&=&\left[(\T+1)^2-5)\right]\left[(\T-1)^2-5)\right]
\\
Q_{2} & = &
\T^4Q_0+144(1+\t^2)-12\T^2(5\t^4+6\t^2+9)
 \\
Q_{3} & = &
6\T\t
\left[
\T^2(\t^4+2\t^2+9)-12(1+\t^2)
\right] .
\end{eqnarray*}%
We note that this is just the second term in the right hand side
of Eq. (\ref{P2e2}) which is responsible for time oscillations.
Therefore if $Q_{1}=0$ the oscillations in the time dependence of
the probability disappear and once again it acquires a monotonous
character. But this time since $Q_1(\T)$ is a bi-quadratic
function, in contrast to the previous case, the  clockwise current
direction probability turns from an oscillating to monotonous
character at two possible values of parameter $\T$,
$\T=\sqrt5\pm1$. In these cases the behavior of probabilities
$P_3^\downarrow(\t)$ and $P_3^\uparrow(\t)$ is illustrated in
Figs. \ref{fig5a}a and \ref{fig5b}b (solid (black lines))
respectively.
We thus observe for $P_3^\downarrow(\t)$ an effect similar to that
described above for $P_1^\downarrow(\t)$ and, in a sense, the
opposite behavior of $P_3^\uparrow(\t)$. The existence of the
critical value is reflected also by the averaged probability
$\overline P_3^\downarrow$ which is plotted in Fig. \ref{fig3b}b
(solid (black) line). It also has a simple analytic expression
\begin{equation}
\overline{P_{3}^{{\downarrow} }}=2\D^2
\frac{13-8\D^2+16\D^4}{(1+4\D^2)^2}
\end{equation}%
with a maximum
$\overline{P_{3}^{{\downarrow} }}\approx 0.91$ at $\D \approx 0.34$.


The next question we study is how the effect, observing for an
idealized closed system, is influenced by a dissipation featuring
open quantum systems \cite{Breuer}. To make rough estimations we
use a phenomenological approach in the density matrix formalism
(see e.g. \cite{Blum}). In this approach a weak coupling of a
system to the environment is described by two parameters,
dephasing $\Gamma _{\varphi }$ and relaxation $\Gamma _{r}$ rates.
Under the initial condition $2\rho(0)=I$ with $I$ being identity
matrix the elements of the density matrix (cf. \cite{ShKOK})
\begin{equation*}
 {\rho }=\frac{1}{2}\left[
\begin{array}{cc}
1+Z & X-iY \\
X+iY & 1-Z%
\end{array}%
\right]
\end{equation*}%
satisfy the
Bloch equations \cite{Blum}
\bea
\dot X& =-2\varepsilon (t)Y-\Gamma _{\varphi }X  \nonumber \\
\dot Y& =-2\D Z+2\varepsilon (t)X-\Gamma _{\varphi }Y  \nonumber \\
\dot Z& =2\D Y-\Gamma _{r}\left( Z-Z(0)\right) .  \nonumber
\eea
Probabilities $P^{{\downarrow}, \uparrow }$ are defined only by
the diagonal entries of the density matrix,
$P^{{\downarrow}, \uparrow }=(1\mp Z(t))/2$.

The relaxation and dephasing effects are shown in Figs.
\ref{fig1}--\ref{fig4} by dot-dashed (red) and double-dot-dashed
(green) lines. As expected, they disturb the system. The influence
of dephasing is more crucial and it should not exceed 5 percent of
$\D$ value for bias (\ref{e3}). We observe an interesting
phenomenon concerning the relaxation. When the probability falls
to zero the relaxation smoothes this behavior and can even revert
it (see green line in Fig. \ref{fig2b}a). Thus, in the current
case the relaxation may be considered as helping to keep the state
of the qubit unchanged.

\section{Concluding remarks}

One of the most popular way to control the qubit state is to drive
a two level system with microwave pulses (see e.g. \cite{Saito})
where, in general, the probability $P^{{\downarrow} }(\tau)$
displays (possibly controllable) Rabi oscillations. On the other
hand it is known that with a specific monochromatic driving force
one can ``freeze'' the state of a two-level system (so called
``coherent destruction of tunnelling'' \cite{Drast}). In this
paper we report on a similar phenomenon but we would like to point
out that in principle it can be realized under other physical
conditions. This possibility is based on the fact that a magnetic
flux with time-dependence like in Eq. (\ref{e3}) may be realized
with the help of a superconducting current. Indeed, according to
resistively shunted junction model \cite{Shmidt} the time
dependence of the current going through the junction, which may
produce the desirable external flux, reads \be\label{Ist}
I_s(t)=I-\frac{I^2-I^2_c}{I+I_c\cos\widetilde\omega t}\,, \qquad
\widetilde\omega=\frac{2e}{\hbar}R\sqrt{I^2-I_c^2} \ee where $R$
is a resistor parallel to the junction, $I$ is an externally
applied constant circuit current and $I_c$ is a critical value of
the persistent current. The time interval, during which the qubit
is in $P_{2}^{{\downarrow} }$ state, depends on frequency $\omega$
from Eq. (\ref{e3}) which according to Eq. (\ref{Ist}) may be
continuously tuned with an externally applied current $I$.

\section*{Acknowledgments}

The authors would like to thank S. Shevchenko for useful
discussions. The work is partially supported by RFBR grant
06-02-16719 and
Russia President grant
871.2008.2. V.V.S. acknowledges a partial
support from INTAS Fellowship Grant for Young Scientists Nr
06-1000016-6264.

\end{document}